\documentclass[12pt]{iopart}
\usepackage[dvips,final]{graphicx}
\usepackage[defs]{ams}
\usepackage{url}
\begin{document}
\title{The role of hydrogen in room-temperature ferromagnetism at graphite surfaces}
\author{H. Ohldag}

\address{Stanford Synchrotron Radiation Lightsource, Stanford University, Menlo Park, CA 94025, USA}
\author{P. Esquinazi}
\address{Institut f\"{u}r Experimentelle Physik II, Universit\"{a}t Leipzig, Linn\'{e}stra{\ss}e 5, 04103 Leipzig, Germany}
\author{E. Arenholz}
\address{Advanced Light Source, Lawrence Berkeley National Laboratory, Berkeley, CA 94720, USA}
\author{D. Spemann, M. Rothermel, A. Setzer, T. Butz}
\address{Institut f\"{u}r Experimentelle Physik II, Universit\"{a}t Leipzig, Linn\'{e}stra{\ss}e 5, 04103 Leipzig,
Germany} 
\begin{abstract}
We present a x-ray dichroism study of graphite surfaces that addresses the origin and magnitude of ferromagnetism in metal-free carbon. We find that, in addition to carbon $\pi$ states, also hydrogen-mediated electronic states  exhibit a net spin polarization with significant magnetic remanence at room
temperature. The observed magnetism is restricted to the top $\approx$10~nm of the irradiated sample where the actual magnetization reaches $ \simeq 15$~emu/g at room temperature. We prove that the ferromagnetism found in metal-free untreated graphite is intrinsic and has a similar origin as the one found in proton bombarded graphite.
\end{abstract}
\submitto{\NJP}
\maketitle 
The possibility of magnetic order in metal-free carbon is fascinating from a fundamental as well as from a technological point of view. In recent years there have been numerous reports of ferromagnetism in virgin graphite \cite{pabloprb02} as well as graphite treated by ion bombardment \cite{pabloprl03,barzola2,xia08} and carbon nanoparticles \cite{yakovprb03,cau06,par08}; it could be shown that the observed magnetism in proton irradiated carbon is not caused by magnetic impurities but is related to the $\pi$-states of carbon \cite{ohldag:07}. However, the question how it is possible to establish ferromagnetic order in carbon without any metallic magnetic or non-magnetic elements still remains unanswered. Several theoretical studies in the past have suggested that absorption of hydrogen at the edges of \cite{kusakabe03} or on \cite{duplock04}  graphene sheets as well as hydrogen chemisorption in graphite \cite{yaz08} may lead to the formation of a spin polarized band at the Fermi level and robust ferromagnetic order. However, there is so far no convincing experimental evidence supporting the influence of hydrogen. While it is obvious that defects or ad-atoms play a central role, the origin of the magnetic moment observed in graphite \cite{pabloprb02} is not yet understood.

Another intriguing question arises from the observation that the apparent magnetization detected in magnetic graphite is typically many orders of magnitude smaller than the one found for ``classical'' magnets like the 3d-transition metals. Apart from the fact that this makes it challenging to obtain a reliable and
detailed understanding of the relevant processes that cause the ferromagnetic order in graphite \cite{jems08}, it also leads to the question how such a system exhibiting a small magnetization and presumably negligible magnetic exchange coupling can be a ferromagnet at room temperature. However, the explanation of such extremely small magnetization resides in the uncertainty of the total ferromagnetic mass in the measured samples and in the role of non-metallic defects. Recent studies on proton- \cite{barzola2} and carbon-irradiated \cite{xia08} graphite managed to provide maximum limits for the induced ferromagnetic mass allowing to estimate magnetization values exceeding $5~$emu/g. Although important evidence has been obtained that supports the role of  vacancies in the graphite ferromagnetism \cite{yan09}, the very origin and extent of surface magnetism \cite{muSR} remains open. For this reason we will investigate the electronic structure of such samples and correlate our findings with the macroscopic magnetic properties.

In this paper we present x-ray absorption (XA) and x-ray magnetic circular dichroism (XMCD) spectra in combination with Superconducting Quantum Interference Device (SQUID) measurements on proton irradiated and non-irradiated highly oriented pyrolytic graphite (HOPG) samples. Soft XA dichroism spectroscopy is an
element specific technique that is sensitive to the magnetic moment of each elemental or chemical species in a complex heterogenous sample. The x-ray energy is chosen such that core level electrons are excited into empty valence states. Using circular polarized x-rays the intensity of the absorption process in a magnetic element depends on the relative orientation between helicity of the x-rays and the magnetic moment of the atom (X-ray magnetic circular dichroism, XMCD)\cite{thole:92}. To obtain an XMCD spectrum two XA spectra with either opposite magnetization or x-ray polarization \cite{sto06} are recorded and compared. It is important to note that XMCD can provide independent information about the magnetic order of different elements or even different chemical states of one element in a sample. Furthermore, it is possible to change the depth sensitivity of the approach by using different detection methods for the absorption yield. We will employ exactly these capabilities of XMCD to show that hydrogen atoms at the surface of graphite play a key role in the ferromagnetism of graphite and that the size of the magnetic moment of graphite at the surface can reach the same order of magnitude than classical ferromagnetic materials.

%
\begin{figure}
\begin{center}
\includegraphics[width=60mm]{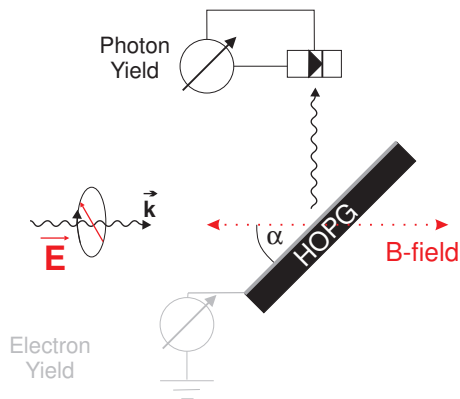}
\caption[]{Experimental geometry. Circular polarized x-rays are incident on the sample under an angle $\alpha$ collinear with the direction of the applied magnetic field. A photodiode at 90$^\circ$ to the incoming x-rays can be used to measure the reflected x-ray intensity at $\alpha=45^\circ$. The absorption yield is measured by monitoring the sample drain current.} \label{fig:Setup}
\end{center}
\end{figure}

The XA experiments were performed using the eight pole magnet \cite{arenholz:05} at the elliptical polarized undulator beamline 4.0.2 at the Advanced Light Source in Berkeley \cite{young:02} at room temperature. The x-ray source provides soft x-rays in the energy range between 250-2000~eV and a typical spectral resolution of $E/\Delta E\approx 5000$ with variable x-ray polarization. The photon energy of the x-ray beam has been calibrated using well known absorption resonances. At the carbon K-resonance the degree of circular polarization of the x-rays emitted by such a device is close to 100\%. The electromagnet provides a magnetic field was applied parallel to the direction of the incoming x-rays while the sample was oriented at an angle $\alpha$ as shown in Fig.~\ref{fig:Setup}. At $\alpha=30^\circ$ only the electron yield (EY) emitted from the sample was recorded, while we also recorded the reflected x-ray intensity (reflection yield, RY) at $\alpha=45^\circ$ using a photodiode. The EY approach provides surface specific information from the first 5-10~nm of the sample \cite{sto06}, while the reflected photons provide more bulk sensitive information (0.1-1$\mu$m) \cite{cxrowww}. XMCD spectra were obtained by switching the direction of the applied field for every data point. The EY and RY data were recorded in an applied field or in remanence after the field was removed.

\begin{table}
\begin{center}
\begin{tabular}{|c|c|c|c|c|c|c|c|c|}
\hline
Sample & Ti   & V    & Cr      & Fe   & Co             & Ni   & Cu   & Zn                \\
\hline
\hline
irr.   & 3.56 & 7.1  & $<$0.03 & 0.25 & $<$0.02        & 0.11 & 0.4  & 0.3               \\
\hline
virgin & 2.69 & 8.52 & 0.11    & 0.61 & $\backsimeq$ 0 & 0.13 & 0.06 & $\lesssim$ 0.06   \\
\hline
\end{tabular}
\caption[]{Metallic impurities in the two HOPG samples
investigated here measured in ppm = $1~\mu$g element per gram
carbon measured by PIXE} \label{tab:impurities}
\end{center}
\end{table}

\begin{figure}
\begin{center}
\includegraphics[width=80mm]{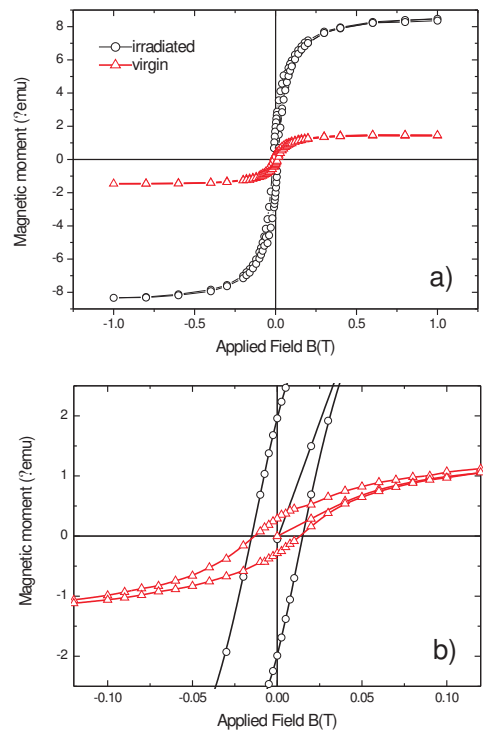}
\caption[]{(a) Hysteresis  loops (magnetic moment $m$ versus applied
field $B$)  measured with SQUID at 300~K for the irradiated (black)
and virgin (red) HOPG samples of identical areas and after
subtraction of a linear diamagnetic background (field applied $\perp
c-$axis)(b) The low-field region of the hysteresis loop is shown to
demonstrate magnetic remanence and coercivity of the two samples. }.
\label{fig:loops}
\end{center}
\end{figure}

The two samples discussed  in this paper were obtained from a HOPG
(0.4$^{\circ}$ rocking curve width) bulk sample. Before performing
the XMCD measurements we confirmed the negligible amount of magnetic
impurities like Fe, Co, Ni by particle induced x-ray emission (PIXE)
and by acquiring XA spectra between 200~eV and 1500~eV. The results
of the very sensitive PIXE measurements are summarized in
table~\ref{tab:impurities}. The total amount of metallic impurities
in both samples is of the order of 10~ppm ($10~\mu$g per gram
carbon), however, the concentration for each of the magnetic elements
Fe, Co and Ni is well below 1~ppm. While one sample, in the following
referred to as virgin sample, was not treated any further, the other
sample was irradiated with a relatively weak fluence of
0.1~nC/$\mu$m$^2$ protons of 2~MeV energy (180~nA ion current).
Hysteresis loops of the measured moments from the two samples are
shown in Fig.~\ref{fig:loops}. Both hysteresis loops show features
that are indicative of ferromagnetic order, like magnetic saturation
at low fields ($\lesssim$0.5\~Tesla), magnetic remanence at zero
field and the existence of a non-vanishing coercive field that is
needed to reverse the magnetization into the opposite direction. To
illustrate all of these three  points we show the overview loops in
figure~\ref{fig:loops}(a) as well as magnified view of the low field
region of the hysteresis loop in panel (b).

\begin{figure}
\begin{center}
\includegraphics[width=80mm]{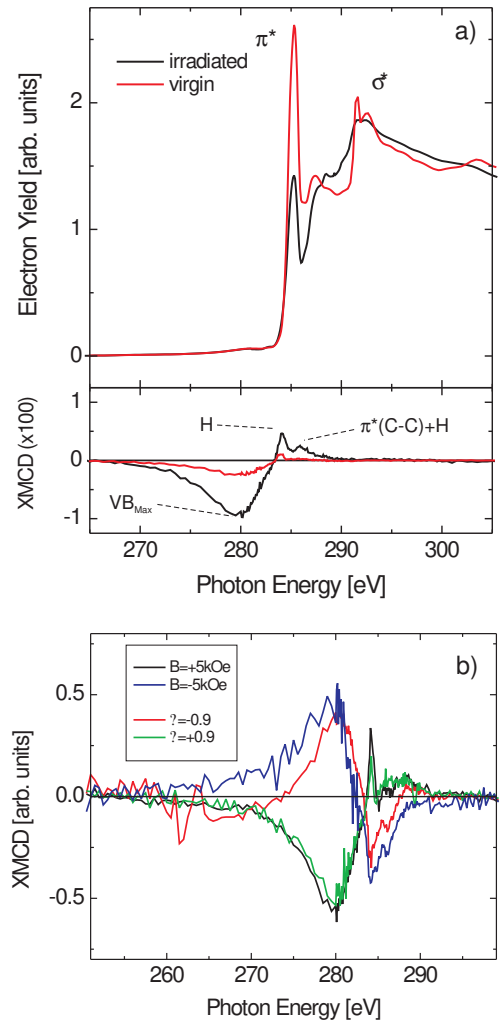}
\caption[]{(a) X-ray absorption spectrum of the  irradiated
(black) and virgin (red) HOPG sample measured using electron yield
as well as the XMCD difference (x100) detected using an applied
field of $\pm 0.5$~Tesla. (b) Four different XMCD spectra measured
by either switching the x-ray polarization or the applied magnetic
field for every data point.}\label{fig:XMCDspectra}
\end{center}
\end{figure}

Carbon K-edge soft x-ray absorption spectra of the virgin and
irradiated HOPG samples  obtained using the electron yield
detection at an angle of incidence $\alpha=30^\circ$ are presented
fig.~\ref{fig:XMCDspectra}(a). The XA spectrum of the virgin
sample shows a narrow resonance at 285.3eV and a broader resonance
around 293eV, which represent the well known excitation of 1s core
level electrons into empty $\pi^\star$ and $\sigma^\star$ bands
\cite{stohrbook}, respectively. The additional feature at the
onset of the $\sigma^\star$ resonance results from an excitonic
transition \cite{bruehwiler:95}. The $\pi^\star$ and
$\sigma^\star$ feature are much broader in the spectra of the
irradiated sample and in particular the intensity of the
$\pi^\star$ resonance is greatly reduced, a sign that the
irradiated sample is more disordered. This observation is
consistent with previous microscopy \cite{ohldag:07} and Raman
\cite{han03} studies, in which we directly compared irradiated and
non-irradiated areas on a carbon sample.

The lower panel in fig.~\ref{fig:XMCDspectra}(a) shows the  XMCD
spectrum obtained in an applied magnetic field of $\pm 0.5$~Tesla
using circular polarized x-rays. To plot the XMCD difference,
which is on average less than 1\% of the XA intensity, on the same
scale as the XA spectrum we multiplied it by a factor of 100. The
excellent signal to noise ratio is achieved by averaging the XMCD
difference over several scans. The onset of the absorption yield
as well as the XMCD intensity  lies well below the
$\pi^\star$-resonance between 270~eV and 275~eV. In our spectra
the Fermi level appears at around 283eV, which means that the XMCD
and XA signal extend to about 10~eV below E$_F$. The appearance of
magnetic circular dichroism in x-ray absorption below the Fermi
level has been previously reported by Mertins {\em et al.}
\cite{mertins04}, who have shown that dichroic effects of
non-magnetic origin can indeed be observed as low as 270~eV in the
Carbon 1s XA spectrum. The broad pre-edge peak of the of the XMCD
spectrum in fig.~\ref{fig:XMCDspectra} exhibits an absolute
maximum at around 3~eV below E$_F$. The energy range for which the
pre-edge XMCD is observed is consistent with the energy spread of
the carbon $\pi$-bands and the maximum of the XMCD at $\simeq
280~$eV corresponds to the valence band maximum $\pi$ valence band
(VB$_\textrm{max}$), see for example \cite{carlisle:00}. The XMCD
spectra of both samples also show a positive peak at 284~eV at the
onset of the $\pi^\star$-resonance close to the Fermi level.
Previous theoretical investigations predicted the occurrence of a
spin polarized band due to H chemisorption
\cite{kusakabe03,duplock04,yaz08} at this energy, which we now
observe in our XA spectrum. To further convince ourselves that the
observed XMCD effects are of magnetic origin we acquired XMCD
spectra by either switching the magnetization on each data point
(constant x-ray polarization) or by switching the x-ray
polarization for each data point (constant field). The four
spectra are shown in fig.~\ref{fig:XMCDspectra}(b). The legend
indicates if either the field or the magnetization has been kept
constant. No averaging has been applied to these spectra, which
explains the higher noise level compared to
fig.~\ref{fig:XMCDspectra}(a). The main features of the XMCD
spectra are reproduced in all four spectra and the sign of the
peaks is reversed upon either switching the polarization or the
magnetization as it is expected for a true intrinsics XMCD effect.

The XMCD spectrum of the irradiated sample shows an additional
feature around 286~eV. X-ray studies on single walled carbon
nanotubes \cite{nikitin:03} and graphite surfaces \cite{nikitin:08}
have shown that C-H bond formation will occur through
re-hybridization of carbon $sp^2$ bonds to $sp^3$ bonds by attachment
of hydrogen, leading to a resonance in the XA spectrum about 1~eV
above the $\pi^\star$ resonance. We find that both, the hydrogen band
as well as the re-hybridized $sp^3$ carbon $\pi$-band exhibit a
magnetic moment in the irradiated sample. The $sp^3$ feature however
is absent in the XMCD spectrum of the virgin sample, while the
feature at 284~eV is again a factor of $\sim$~4 smaller. Combined
these observations indicate that the effect of proton irradiation is
the incorporation of hydrogen through re-hybridization. This may
initially be advantageous for the evolution and stabilization of
ferromagnetism in graphite. However, it also becomes evident that
intense proton irradiation in particular at ambient temperatures will
significantly enhance the diffusion of H and also disturb the
crystallographic and hence the electronic structure of graphite
reducing the magnetic order. We note that no XMCD intensity is
observed at photon energies correlated with the presence of surface
oxides around 289eV, excluding the possibility that surface oxidation
plays a significant role in the ferromagnetism of graphite.

\begin{figure}
\begin{center}
\includegraphics[width=80mm]{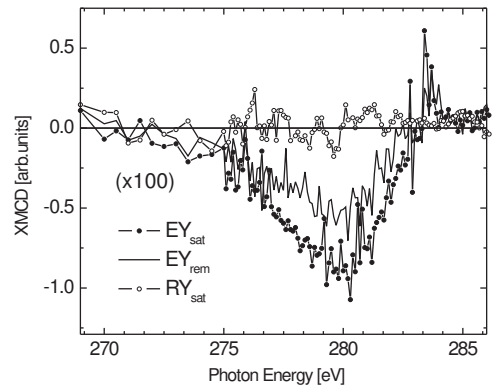}
\caption[]{XMCD spectra (x100) of the irradiated HOPG samples obtained using EY in an applied field of $\pm$0.7~T ($\bullet$) or in remanence after saturating the sample in a field of $\pm$0.7~T ($-$). Also shown is the XMCD spectrum acquired in saturation using the reflection yield ($\circ$). Lines are added as a guide to the eye. All spectra were obtained at $\alpha=45^\circ$.} \label{fig:XMCDbulksurfrem}
\end{center}
\end{figure}

Figure~\ref{fig:XMCDbulksurfrem} shows three XMCD spectra (no averaging as in fig.~\ref{fig:XMCDspectra}(a)) obtained from the irradiated sample. The spectra were acquired at an angle of incidence $\alpha=45^\circ$, so that the intensity of the RY could be detected at the same time as the EY in our setup. The RY provides bulk sensitive information while the EY is a surface sensitive approach. This is in particular true at energies below the $\pi$-resonance where the absorption cross section is small and the x-rays can penetrate deeper into the material, while the escape depth of the secondary electrons contributing to the EY remains small. The XMCD signal acquired using EY shows the prominent feature at 280~eV, which clearly persists after the external field is removed, showing a clear magnetic remanence of about 60\% of the saturation value. On the other hand, there is no XMCD detected in the RY spectrum in fig~\ref{fig:XMCDbulksurfrem}. Taking into account the different probing depths of the two approaches (10~nm for EY versus 1~$\mu$m in RY) we conclude that the observed ferromagnetism  in the irradiated sample originates predominantly from the top 10~nm of the sample. Note, that this does not imply that the ferromagnetic order vanishes completely in the bulk of the sample, instead it means that the (orbital) magnetic moment per atom is smaller than our sensitivity of $10^{-3}\mu_B$.

\begin{figure}
\begin{center}
\includegraphics[width=80mm]{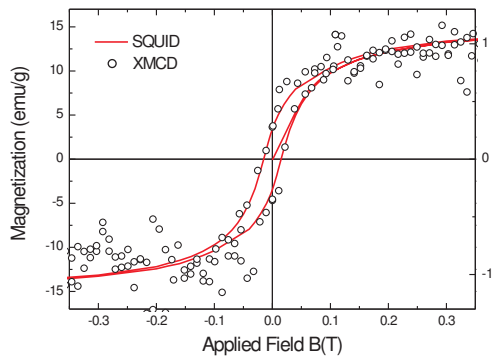}
\caption[]{Hysteresis loops of the irradiated sample acquired using SQUID ($-$) or XMCD ($\circ$). The axis on the left side show the magnetization of the sample in emu/g. The axis of the right for the XMCD loop is chosen such that the saturation values coincide for comparison.} \label{fig:XMCDSQUIDloop}
\end{center}
\end{figure}

The information that the majority of the magnetic moment originates from the surface allows us now to estimate the true magnetization values of the HOPG surface by normalizing the magnetic moment obtained by the SQUID hysteresis loop in Fig.~\ref{fig:XMCDspectra}(a) with the correct thickness. For simplicity we assume a homogeneous distribution of 90\% of the magnetic moment within the first 10~nm and we find that the average saturation magnetization for the irradiated sample is of the order of $15$~emu/g. Fig.~\ref{fig:XMCDSQUIDloop} shows the corrected SQUID hysteresis loop displaying the magnetization of the sample surface. For comparison the magnetic moment of Fe is about 220~emu/g or Ni 55~emu/g, which means that the magnetization at the surface of HOPG may reach up to 25\% of that of Ni, or on average about 0.1$~\mu_B$ per C-atom. Note that a direct determination of the total magnetic moment (spin and orbit) of the carbon atoms using x-ray absorption sum rules is not possible because the carbon K-edge XMCD only probes the orbital moment. Nevertheless, both techniques (SQUID and XMCD) measure the same macroscopic magnetic properties. For comparison we show a hysteresis loop obtained using XMCD at a photon energy of 280eV in fig.\ref{fig:XMCDSQUIDloop} which nicely follows the SQUID hysteresis loop.

In conclusion we report the  observation of ferromagnetic order at surfaces of metal free HOPG before and after proton irradiation. The results in virgin HOPG clearly indicate that the observed ferromagnetism in untreated, pure graphite \cite{pabloprb02} is intrinsic. We find that the observed XMCD signal originates mostly from the near surface region ($\approx$ 10~nm) of the sample where the saturation magnetization may reach up to 25\% of that of Ni. The rather large magnetic moment of ferromagnetic HOPG is now consistent with the occurrence of room temperature ferromagnetism. Furthermore, it points toward the possibility that the magnetic properties of carbon can indeed play a significant role in bio-compatible nano-applications as well as an organic magnetic ``substrate" in
chiral selective chemistry \cite{ros08}. The XMCD line shape shows that chemisorbed hydrogen and C-H bond states as well as carbon $\pi$-states exhibit a net spin polarization. While the proton irradiation initially increases the magnetic moment by formation of C-H bonds it also disturbs the system for the same reason by introducing disorder and formation of $sp^3$-bonds providing an explanation on the limit to which the magnetization in carbon can be increased by ion irradiation. Our results also indicate that hydrogen is not implanted -- the implantation probability of 2~MeV photons is very small at the surface -- but should come from dissociation of H$_2$ molecules by the proton irradiation, since a large hydrogen concentration is always present in the near surface region \cite{rei06}.

H.O. would like to thank  Joachim St\"ohr and Hans-Christoph Siegmann for their support and stimulating  discussions. SSRL and ALS are national user facilities supported by the Department of Energy, Office of Basic Energy Sciences. SSRL is operated by Stanford University and ALS is operated by the University of California under contract No. DE-AC02-05CH11231. The work at the University of Leipzig is supported by the DFG under DFG ES 86/16-1 and the European Union
project ``Ferrocarbon".


\bibliographystyle{unsrt}

\begin{thebibliography}{10}

\bibitem{pabloprb02}
P.~Esquinazi, A.~Setzer, R.~H\"ohne, C.~Semmelhack, Y.~Kopelevich, D.~Spemann,
  T.~Butz, B.~Kohlstrunk, and M.~L\"osche.
\newblock Ferromagnetism in oriented graphite samples.
\newblock {\em Phys. Rev. B}, 66:024429--1--10, 2002.

\bibitem{pabloprl03}
P.~Esquinazi, D.~Spemann, R.~H\"ohne, A.~Setzer, K.-H. Han, and T.~Butz.
\newblock Induced magnetic ordering by proton irradiation in graphite.
\newblock {\em Phys. Rev. Lett.}, 91:227201--1--4, 2003.

\bibitem{barzola2}
J.~Barzola-Quiquia, P.~Esquinazi, M.~Rothermel, D.~Spemann, T.~Butz, and
  N.~Garc\'ia.
\newblock Experimental evidence for two-dimensional magnetic order in proton
  bombarded graphite.
\newblock {\em Phys. Rev. B}, 76:161403{\rm (R)}, 2007.

\bibitem{xia08}
H.~Xia, W.~Li, Y.~Song, X.~Yang, X.~Liu, M.~Zhao, Y.~Xia, C.~Song, T.-W. Wang,
  D.~Zhu, J.~Gong, and Z.~Zhu.
\newblock Tunable magnetism in carbon-ion-implanted highly oriented pyrolytic
  graphite.
\newblock {\em Adv. Mater.}, 20:1--5, 2008.

\bibitem{yakovprb03}
Y.~Kopelevich, R.~R. da~Silva, J.~H.~S. Torres, A.~Penicaud, and T.~Kyotani.
\newblock Local ferromagnetism in microporous carbon with the structural
  regularity of zeolite {Y}.
\newblock {\em Phys. Rev. B}, 68:092408--1--4, 2003.

\bibitem{cau06}
R.~Caudillo, X.~Gao, R.~Escudero, M.~Jos\'e-Yacaman, and J.~B. Goodenough.
\newblock Ferromagnetic behavior of carbon nanospheres encapsulating silver
  nanoparticles.
\newblock {\em Phys. Rev. B}, 74:214418, 2006.

\bibitem{par08}
N.~Parkanskya, B.~Alterkopa, R.~L. Boxmana, G.~Leitusb, O.~Berkhc, Z.~Barkayd,
  Yu. Rosenberg, and N.~Eliaz.
\newblock Magnetic properties of carbon nano-particles produced by a pulsed arc
  submerged in ethanol.
\newblock {\em Carbon}, 46:215--219, 2008.

\bibitem{ohldag:07}
H.~Ohldag, T.~Tylisczak, R.~{H\"{o}hne}, D.~Spemann, P.~Esquinazi,
  M.~Ungureanu, and T.~Butz.
\newblock {\em Phys. Rev. Lett.}, 98:187204, 2007.

\bibitem{kusakabe03}
K.~Kusakabe and M.~Maruyama.
\newblock Magnetic nanographite.
\newblock {\em Phys. Rev. B}, 67:092406--1--4, 2003.

\bibitem{duplock04}
E.~J. Duplock, M.~Scheffler, and P.~J.~D. Lindan.
\newblock Hallmark of perfect graphite.
\newblock {\em Phys. Rev. Lett.}, 92:225502--1--4, 2004.

\bibitem{yaz08}
O.~V. Yazyev.
\newblock Magnetism in disordered graphene and irradiated graphite.
\newblock {\em Phys. Rev. Lett.}, 101:037203, 2008.

\bibitem{jems08}
P.~Esquinazi, J.~Barzola-Quiquia, D.~Spemann, M.~Rothermel, H.~Ohldag,
  N.~Garc\'ia, A.~Setzer, and T.~Butz.
\newblock Magnetic order in graphite: Experimental evidence, intrinsic and
  extrinsic difficulties.
\newblock {\rm J. Magn. Magn. Mat. (to be published), see {\rm
  arXiv:0902.1671}}, 2008.

\bibitem{yan09}
X.~Yanga, H.~Xiab, X.~Qinc, W.~Lia, Y.~Daia, X.~Liua, M.~Zhaoa, Y.~Xiaa,
  S.~Yana, and B.~Wangc.
\newblock Correlation between the vacancy defects and ferromagnetism in
  graphite.
\newblock {\em Carbon}, 47:1399--1406, 2009.

\bibitem{muSR}
M.~Dubman, T.~Shiroka, H.~Luetkens, M.~Rothermel, F.~J. Litterst, E.~Morenzoni,
  A.~Suter, D.~Spemann, P.~Esquinazi, A.~Setzer, and T.~Butz.

\bibitem{thole:92}
B.~T. Thole, P.~Carra, F.Sette, and G.van der Laan.
\newblock {\em Phys. Rev. Lett.}, 68:1943, 1992.

\bibitem{sto06}
J.~Stoehr and H.~C. Siegmann.
\newblock {\em Magnetism - From Fundamentals to Nanoscale Dynamics}, volume 152
  of {\em Springer Series in Solid State Sciences}.
\newblock Springer Heidelberg, 2006.

\bibitem{arenholz:05}
E.~Arenholz and S.~Prestomon.
\newblock {\em Rev. Sci. Instrum.}, 76:083908, 2005.

\bibitem{young:02}
A.~Young, E.~Arenholz, S.~Marks, R.~Schlueter, C.~Steier, H.~Padmore,
  A.~Hitchcock, and D.~Castner.
\newblock {\em Jour. Synch. Rad.}, 9:270, 2002.

\bibitem{cxrowww}
See for example: \url{http://henke.lbl.gov/optical_constants/atten2.html},.

\bibitem{stohrbook}
J.~Stoehr.
\newblock {\em NEXAFS Spectroscopy}, volume~25 of {\em Springer Series in
  Surface Science}.
\newblock Springer, Heidelberg, 1992.

\bibitem{bruehwiler:95}
P.~Br\"uhwiler, A.~Maxwell, C.~Puglia, A.~Nilsson, S.~Andreson, and
  N.~M{\aa}rtenson.
\newblock {\em Phys. Rev. Lett.}, 74:614, 1995.

\bibitem{han03}
K.-H. Han, D.~Spemann, P.~Esquinazi, R.~H\"ohne, V.~Riede, and T.~Butz.
\newblock Ferromagnetic spots in graphite produced by proton irradiation.
\newblock {\em Adv. Mater.}, 15:1719--1722, 2003.

\bibitem{mertins04}
H.-Ch. Mertins, S.~Valencia, W.~Gudat, P.~M. Oppeneer, O.~Zaharko, and
  H.~Grimmer.
\newblock Direct observation of local ferromagnetism on carbon in {C/Fe}
  multilayers.
\newblock {\em Europhys. Lett.}, 66:743--748, 2004.

\bibitem{carlisle:00}
J.~Carlisle, S.~Blankenshop, L.~Termminello, J.~Jia, T.~Callcott, D.~Ederer,
  R.~Perera, and F.~Himpsel.
\newblock {\em J. Electron Spectrosc. Rel. Phenom.}, 110:323, 2000.

\bibitem{nikitin:03}
A.~Nikitin, H.~Ogasawara, D.~Mann, R.~Denecke, Z.~Zhang, H.~Dai, K.~Cho, and
  A.~Nilsson.
\newblock {\em Phys. Rev. Lett.}, 95:225507, 2003.

\bibitem{nikitin:08}
A.~Nikitin, L.-A. N{\aa}slund, Z.~Zhang, and A.~Nilsson.
\newblock {\em Surf. Sci.}, 602:2575, 2008.

\bibitem{ros08}
R.~A. Rosenberg, M.~Abu Haija, and P.~J. Ryan.
\newblock {\em Phys. Rev. Lett.}, 101:178301, 2008.

\bibitem{rei06}
P.~Reichart, D.~Spemann, A.~Hauptner, A.~Bergmaier, V.~Hable, R.~Hertenberger,
  C.~Greubel, A.~Setzer, T.~Butz, G.~Dollinger, D.N. Jamieson, and
  P.~Esquinazi.
\newblock {\em Nucl. Instrum. Methods Phys. Res. B}, 249:286, 2006.

\end{thebibliography}

\end{document}